\RequirePackage{fix-cm}
\documentclass[twocolumn,epjc3]{svjour3}  

\smartqed  % flush right qed marks, e.g. at end of proof
\RequirePackage{graphicx}

\usepackage{hyperref}
\usepackage{color}
\usepackage{amsmath}

\usepackage{lineno}
\journalname{Eur. Phys. J. C}
\begin{document}
\newcommand{\nuc}[2]{$^{#2}\rm #1$}

\newcommand{\bb}[1]{$\rm #1\nu \beta \beta$}
\newcommand{\bbm}[1]{$\rm #1\nu \beta^- \beta^-$}
\newcommand{\bbp}[1]{$\rm #1\nu \beta^+ \beta^+$}
\newcommand{\bbe}[1]{$\rm #1\nu \epsilon \epsilon$}
\newcommand{\bbep}[1]{$\rm #1\nu \rm EC \beta^+$}

\newcommand{\pic}[5]{
       \begin{figure}[ht]
       \begin{center}
       \includegraphics[width=#2\textwidth, keepaspectratio, #3]{#1}
       \end{center}
       \caption{#5}
       \label{#4}
       \end{figure}
}

\newcommand{\apic}[5]{
       \begin{figure}[H]
       \begin{center}
       \includegraphics[width=#2\textwidth, keepaspectratio, #3]{#1}
       \end{center}
       \caption{#5}
       \label{#4}
       \end{figure}
}

\newcommand{\sapic}[5]{
       \begin{figure}[P]
       \begin{center}
       \includegraphics[width=#2\textwidth, keepaspectratio, #3]{#1}
       \end{center}
       \caption{#5}
       \label{#4}
       \end{figure}
}

\newcommand{\picwrap}[9]{
       \begin{wrapfigure}{#5}{#6}
       \vspace{#7}
       \begin{center}
       \includegraphics[width=#2\textwidth, keepaspectratio, #3]{#1}
       \end{center}
       \caption{#9}
       \label{#4}
       \vspace{#8}
       \end{wrapfigure}
}

\newcommand{\baseT}[2]{\mbox{$#1\times10^{#2}$}}
\newcommand{\baseTsolo}[1]{$10^{#1}$}
\newcommand{\THL}{$T_{\nicefrac{1}{2}}$}

\newcommand{\UBI}{$\rm cts/(kg \cdot yr \cdot keV)$}

\newcommand{\Uflux}{$\rm m^{-2} s^{-1}$}
\newcommand{\Ucpd}{$\rm cts/(kg \cdot d)$}
\newcommand{\Uexpo}{$\rm kg \cdot d$}

\newcommand{\Qbb}{$\rm Q_{\beta\beta}\ $}

\newcommand{\validate}{\textcolor{blue}{\textit{(validate!!!)}}}

\newcommand{\improve}{\textcolor{blue}{\textit{(improve!!!)}}}

\newcommand{\missing}[1]{\textcolor{red}{\textbf{...!!!...} #1}\ }

\newcommand{\missref}{\textcolor{red}{[reference!!!]}\ }

\newcommand{\quanta}{\textcolor{red}{\textit{(quantitativ?) }}}

\newcommand{\misscite}{\textcolor{red}{[citation!!!]}}

%K42
\newcommand{\PC}{$N_{\rm peak}$}
\newcommand{\BIC}{$N_{\rm BI}$}
\newcommand{\PAPR}{$R_{\rm p/>p}$}

\newcommand{\PCR}{$R_{\rm peak}$}

%Pd

\newcommand{\gline}{$\gamma$-line}
\newcommand{\glines}{$\gamma$-lines}
\newcommand{\gray}{$\gamma$-ray}
\newcommand{\grays}{$\gamma$-rays}

%general

\newcommand{\tab}{Tab.~}
\newcommand{\eq}{Eq.~}
\newcommand{\fig}{Fig.~}
\renewcommand{\sec}{Sec.~}
\newcommand{\chap}{Chap.~}

 \newcommand{\fn}{\iffalse \fi} %footnote explaination
 \newcommand{\tx}{\iffalse \fi} %text explaination
 \newcommand{\txe}{\iffalse \fi} %text extended explaination
 \newcommand{\sr}{\iffalse \fi} %section reference explaination

\title{Search for resonant neutrinoless double electron capture in $^{152}$Gd and other rare decays in Gd isotopes}


\author{M. Laubenstein\thanksref{e1,addr1}
        \and
        B. Lehnert\thanksref{e2,addr2} %etc.
        \and
       S. S. Nagorny\thanksref{e3,addr3} %etc.
	\and
	S. Nisi\thanksref{e4,addr1}
}

\thankstext{e1}{e-mail: matthias.laubenstein@lngs.infn.it}
\thankstext{e2}{e-mail: bjoernlehnert@lbl.gov}
\thankstext{e3}{e-mail: sn65@queensu.ca}
\thankstext{e4}{e-mail: stefano.nisi@lngs.infn.it}


\institute{INFN - Laboratori Nazionali del Gran Sasso, 67100 Assergi (AQ), Italy \label{addr1}
           \and
           Nuclear Science Division, Lawrence Berkeley National Laboratory, Berkeley, CA 94720, U.S.A. \label{addr2}
           \and
           Queen's University, Physics Department, Kingston, ON, K7L 3N6, Canada \label{addr3}
}

\date{Received: date / Accepted: date}

% add and check reference
% check detector setup
%
%
%
%
%

\maketitle

\begin{abstract}
A search for rare decays of gadolinium isotopes was performed with an ultra-low background high-purity germanium detector at Gran Sasso Underground Laboratory (Italy). A 198~g Gd$_2$O$_3$ powder sample was measured for 63.8~d with a total Gd exposure of 12.6~kg$\times$d. 
$^{152}$Gd is the most promising isotope for resonant neutrinoless double electron capture which could significantly enhance the decay rate over other neutrinoless double beta decay processes. The half-life for this decay was constrained to $>4.2\times10^{12}$~yr (90\% credibility). This limit is still orders of magnitude away from theoretical predictions but it is the first established limit on the transition paving the way for future experiments. 
In addition, other rare alpha and double beta decay modes were investigated in $^{152}$Gd, $^{154}$Gd, and $^{160}$Gd with half-life limits in the range of $10^{17-20}$~yr.

\end{abstract}

%%%%%%%%%%%%%%%%%%%%%%%%%%%%%%%%%%%%%%%%%%%%%%%%%%%%%%%%%%%%%
%%%%%%%%%%%%%%%%%%%%%%%%%%%%%%%%%%%%%%%%%%%%%%%%%%%%%%%%%%%%%
%%%%%%%%%%%%%%%%%%%%%%%%%%%%%%%%%%%%%%%%%%%%%%%%%%%%%%%%%%%%%
\section{Introduction}
\label{intro}

%- DBD\\

Neutrinoless double beta (\bb{0}) decay is a process that violates lepton number and is one of the most promising searches for physics beyond the Standard Model. Its observation would imply the Majorana nature of neutrinos and could lead to an explanation for the matter-antimatter asymmetry in the Universe (see e.g.\ Ref.~\cite{dep18}).\\

This second-order weak nuclear process is intensively investigated for $\beta^-\beta^-$ decays on the neutron-rich side of the nuclide chart:
\begin{equation}
0\nu\beta^-\beta^-: \hspace{3.5pc} (Z, A) \longrightarrow (Z + 2, A) + 2 \, e^- \,
\end{equation}

The signature is a mono-energetic peak of the two electrons at the Q-value of the decay. See Ref.~\cite{DBDreview} for a review. 
The process can also occur on the proton-rich side of the nuclide chart through double electron capture ($\epsilon$), $\beta^+$ decay, or combinations of these:
\begin{align}
&0\nu\epsilon \epsilon:& 2 e^- + (Z,A) &\longrightarrow (Z-2,A)   \\
&0\nu\epsilon\beta^+:&  e^- + (Z,A) &\longrightarrow (Z-2,A) + e^+  \\
&0\nu\beta^+\beta^+:&  (Z,A) &\longrightarrow (Z-2,A) + 2 e^+   
\end{align}

The decay modes containing positrons in the final state reduce the kinematic phase space by two times 511~keV for each $e^+$ and make these decays less likely. On the other hand, the annihilation of positrons can create an enhanced experimental signature. 
In either way, the lepton number violating process would share the variety of decay modes and has lower expected rates. This makes proton-rich double beta decay isotopes in general less attractive for searches compared to $\beta^-\beta^-$ decay isotopes with a more unique signature.

%- resonant 0nEC

An interesting exception is \bbe{0} modes which formally only have low energy particles from the atomic shell restructuring of the two electron captures in the final state, i.e.\ x-rays or Auger electrons. 
The remaining energy can be released by a Bremsstrahlung photon \cite{Doi93}. 
Other processes are two Bremsstrahlung photons or an $e^-$-$e^+$ pair in case sufficient decay energy is available. However, those processes have a more complex experimental signature and are additionally suppressed by additional vertices in the interaction. 

The lack of final state particles opens the possibility of a direct transition between initial and final state nuclei with resonance enhancement if the system is degenerate in energy with the ground state or an excited state in the daughter. 
The closer the initial and final state energies are, the stronger is the resonance enhancement. Out of 34 double beta decay candidate isotopes on the proton-rich side, only a handful have a suitable nuclear system with degeneration on the keV scale. Exact Q-values and atomic masses for these candidates were recently re-measured using state-of-the-art Penning trap setups. Eliseev et al.\ \cite{Eliseev11} found that \nuc{Gd}{152} is, in fact, the best isotope with a Q-value of $Q_{\epsilon\epsilon}= 55.70(18)$~keV and a mass difference $\Delta = Q_{\epsilon\epsilon} - E = 0.91(18)$~keV. They predict a possible resonance enhancement factor of \baseT{6}{6} compared to the reference of \nuc{Fe}{54} which is not resonant enhanced.

The lack of final state particles in the \bbe{0} ground state transition, leaves the atomic shell restructuring as the only experimental signature. For HPGe \gray\ spectroscopy, as used in this work, we consider only x-ray emissions. Since the detection efficiency is steeply decreasing with lower energy, we focus on the x-rays with the highest energies.

The two captures occur from electron shells with binding energies $E_{\epsilon 1}$ and $E_{\epsilon 2}$. 
Typically, K and L shell captures are the most likely. In $0^+ - 0^+$ transitions two K-shell captures are spin-suppressed making K+L captures the most likely \cite{Doi93}. 
Here we focus specifically on the K+L$_1$ capture case since this transition is expected to have the highest resonance enhancement for \nuc{Gd}{152} \cite{Eliseev11}. 
Captures from less bound electron shells (e.g.\ L+L) are also possible but neglected in this search. Less bound shells have less overlap with the nucleus, have reduced capture probability, and are not resonance enhanced in \nuc{Gd}{152}. 
 
In the daughter element samarium, K-shell and L-shell binding energies are 46.849(13)~keV and 7.74793(72)~keV, respectively \cite{NuclDataX}.
Filling the hole in the K shell yields the highest energetic x-rays of 39 to 47~keV. Filling the hole in the L$_1$ shell results in x-rays between 5 and 8~keV, which is not detectable with the setup used in this study. The possible transitions and x-ray energies in samarium are shown in \tab \ref{tab:capture_xrays}.
Only a subset of theoretical x-ray transitions (second column) has been experimentally observed (third column). 
To obtain the emission probabilities for a given x-ray energy per double electron capture of \nuc{Gd}{152} to \nuc{Sm}{152} decay, we use a proxy decay: \nuc{Eu}{150} to \nuc{Sm}{150} with a 100\% decay branch of single electron capture and the same atomic shell configuration of the daughter. The measured x-ray emissions and their probabilities are shown in the forth and fifth column of \tab \ref{tab:capture_xrays}.

\begin{table}[t]
\begin{center}
\begin{tabular}{lll|lr}
\hline
Trans. & E$_{\rm th}$  &  E$_{\rm exp}$  & E$_{\rm sig}$  &  p$_{\rm sig}$ \\
 & [keV] &  [keV]  & [keV]  &  \\
\hline
KL$_1$	  & 39 101.2(16)  & 	             -     	         &      	\\
KL$_2$	  & 39 524.3(12)  & 	   39 523.39(10)    	 &      39.5    & 22.7\%    \\
KL$_3$	  & 40 119.4(11)  & 	   40 118.481(60)    	 &      40.1    & 40.8\%   \\
KM$_1$	  & 45 110.3(17)  & 	              -    	          &     	\\
KM$_2$	  & 45 293.9(20)  & 	   45 288.6(49)    	  &     45.3  & 4.1\%      \\
KM$_3$	  & 45 418.1(19)  & 	   45 413.0(49)    	  &     45.4  & 7.9\%        \\
KM$_4$	  & 45 728.1(18)  & 	   45 731.4(75)    	  &             \\ 
KM$_5$	  & 45 756.9(16)  & 	   45 731.4(75)   	  &             \\
KN$_1$	  & 46 488.6(44)  & 	             -     	           &    	\\
KN$_2$	  & 46 530(16)     & 	   46 575(26) 	           &    46.6 & 2.6\%	\\
KN$_3$	  & 46 588.2(15)  & 	   46 575(26)    	           &    	\\
KN$_4$	  & 46 709.1(38)  & 	          -        	            &   	\\
KN$_5$	  & 46 706.4(11)  & 	          -        	           &    	\\
\hline
L$_n$X$_y$ & 5.0 - 7.7       &            5.0 - 7.7              &          & 14.0\%    \\
other             & 			& 					& 	      &  7.9\%    \\
\hline
\end{tabular}
\medskip
\caption{\label{tab:capture_xrays} Approximation of x-ray signature from known information in samarium. The first column shows the electron shell transitions. The second and third columns show theoretical and experimental x-ray energies according to \cite{NuclDataX}. The fourth and fifth columns show the measured x-ray energies and emission probabilities for the electron capture of \nuc{Eu}{150} to \nuc{Sm}{150}, used as a proxy for \nuc{Gd}{152} to \nuc{Sm}{152}. The combined x-ray emissions in the last two columns are used to construct the experimental signature occurring in 78.1\% of all decays. 
Transitions to the L-shell occur in 14\% of all cases. Other transitions such as internal conversion make up 7.9\% of the signature.
}
 \end{center}
\end{table}

Hence, we approximate the signature of the x-ray emission of the K+L$_1$ double electron capture as two independent electron captures from these shells.  An obvious difference between single-electron capture and double-electron capture is that an additional hole is present in the shell. Explicitly, this makes the transition from L$_1$ to K less likely, since only one electron remains in the L$_1$ state. However, the x-ray from this specific transition is not observed in literature for the proxy decay of \nuc{Eu}{150} to \nuc{Sm}{150}. 
%Hence, the measured emission probabilities of the other x-rays are independent of its existence \missing{this is probably not correct?!!!}.   
Hence, the difference between single and double EC is neglected here.
In addition, we assume that the molecular form of Gd$_2$O$_3$ does not affect the double electron capture probabilities nor the x-ray emission energies and probabilities\footnote{Gadolinium has fully filled K, L, and M shells in the atom and partially filled N, O, and P shells. Only the partially filled shells are redistributed in chemical compounds and could slightly change the atomic relaxation after electron capture. The KN$_i$ transitions only contribute 2.6\% to the experimental signature in this search and any small changes due to chemical bindings are negligible for the results.}

\begin{figure}
  \centering
  \includegraphics[width=0.5\textwidth]{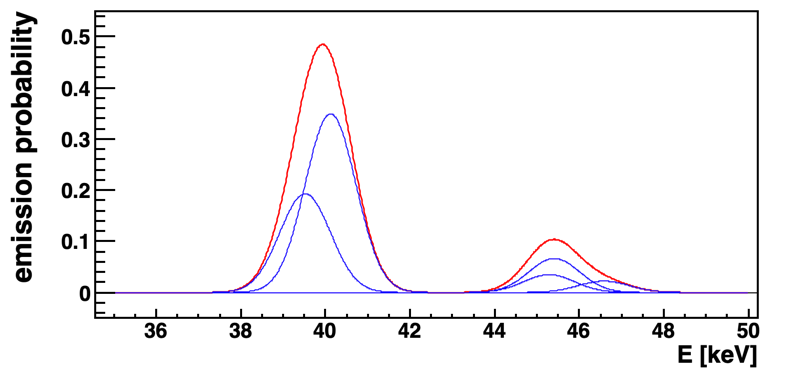}\\
  \includegraphics[width=0.5\textwidth]{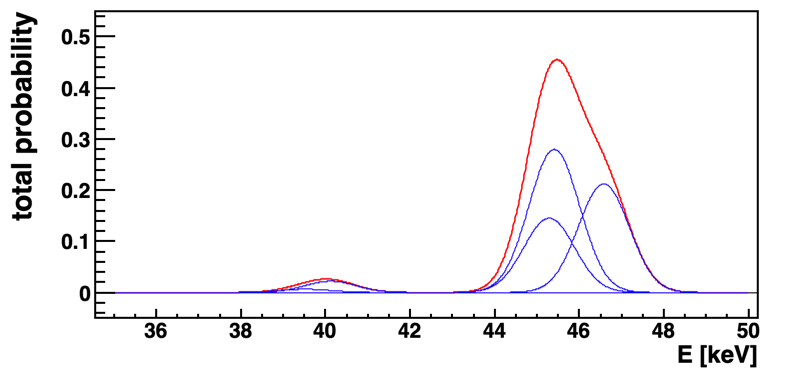}
  %\vspace{5cm}
\caption{Experimental signature of \bbe{0} in \nuc{Gd}{152}. The blue curves show the individual x-ray lines folded with the detector resolution. The red curves shows the sum of the signatures. Top: Emission probability of each x-ray line. Bottom: Emission and detection efficiency of each x-ray line. The strong energy dependence of the detection efficiency is significantly changing the signature in this energy region. The bottom p.d.f.\ is used in the analysis. Note that each x-ray line has an individual probability uncertainty and is allow to vary within its prior constraints.  }
  \label{pic:expSignature}
\end{figure}

The experimental signature of \nuc{Gd}{152} \bbe{0} in search is shown in the top panel of \fig \ref{pic:expSignature}. It contains five x-rays between 39 and 47~keV which sum up to a total emission probability of 78.1\% per decay. In the remaining 21.9\% of cases, the shell-restructuring proceeds in ways not detectable by the experimental setup.
Also shown in \fig \ref{pic:expSignature} (bottom panel) are the x-ray peaks taking into account the detection efficiency. The steep decrease of detection efficiency between 35 and 50 keV shifts the most prominent signal region to about 46~keV.

To our knowledge, this is the first time the atomic shell signature from double electron capture is explicitly decomposed for \gray\ spectroscopy in a source$\neq$detector experimental approach. 
Previous \bbe{2} observations in \nuc{Xe}{124} within the Xenon dark matter experiment are based on calorimetric source$=$detector measurements which are sensitive to the total binding energy.
The measured half-life of \baseT{1.8}{22}~yr \cite{apr19} is the first compelling evidence of \bbe{2} and the slowest nuclear decay ever observed.

The study of other rare nuclear decays in Gd isotopes is also possible. This research can provide experimental information that deepens the understanding of nuclear structure, and has applications in other fields such as nuclear chronometry. Additionally, the information gained can be relevant for long-lived background sources in other rare event searches.

In this work we pursued a generic approach, investigating rare alpha and $0\nu/2\nu\beta\beta$ decays in gadolinium isotopes similar to the study in \cite{Laubenstein22}.
The two-neutrino double beta (\bb{2}) decay mode is a second-order weak process allowed within the standard model. It has been experimentally observed in 11 isotopes with half-lives in the range of \baseTsolo{18-21}~yr using the ``source$\, =\, $detector'' approach. Decays into excited states of the daughter isotope have been observed in \nuc{Nd}{150} and \nuc{Mo}{100} \cite{ESAverage} using the ``source$\, \neq\, $detector'' approach and HPGe detectors.

In this work, we study \nuc{Gd}{160} which can double beta decay into various excited states. 
In addition, we extend the experimental information on rare Gd $\alpha$-decays into excited states. 
Gadolinium contains four isotopes, \nuc{Gd}{152}, \nuc{Gd}{153}, \nuc{Gd}{154}, and \nuc{Gd}{155}, which can undergo $\alpha$-decay. 
\nuc{Gd}{153} has a short half-life of $T_{1/2} = 241.6$~d (EC) and is not naturally occurring in the sample. \nuc{Gd}{155} has a low Q-value of $81.5 \pm 0.7$~keV and an expected half-life $>10^{300}$~yr. Hence, we limit our search to \nuc{Gd}{152} and \nuc{Gd}{154}.

\begin{table*}[t]
\begin{center}
\begin{tabular}{llllllll}
\hline
isotope &  abundance & daughter & Q-value & mode &  level   & T$_{1/2}^{\rm th}$ & T$_{1/2}^{\rm exp}$ previous   \\
             &        [\%]          &            &  [keV]    &            & $J^\pi$ [keV]   &     [yr]                    &  [yr]             \\
\hline
\nuc{Gd}{152} & 0.20           & \nuc{Sm}{148}  &  2204.9           &  $\alpha$                              &  $2^+_1$ 550.3 &      \baseT{2}{25} ($^1$)&   ---       \\
                      &                     & \nuc{Sm}{152}  &  55.70(18)      & $0\nu\epsilon\epsilon$  &  $2^+_1$ 0         &  \baseT{6.8}{27}-\baseT{3.8}{30}  \cite{0neeReview} &   \baseT{>6.0}{8} ($^2$)  \cite{Nozzoli18}     \\

\nuc{Gd}{154} & 2.18 & \nuc{Sm}{150}  &  919.2           &  $\alpha$  &  $2^+_1$ 334.0 &      \baseT{5}{80} ($^1$) &  ---             \\

\nuc{Gd}{160} & 21.86       &  \nuc{Dy}{160} & 1731.0         &  $0\nu\beta\beta$  &  $2^+_1$ 86.8  &      --- & \baseT{>1.3}{21}   \cite{Danevich01}           \\
		      &                  &                        & 		          &  $2\nu\beta\beta$  &  $2^+_1$ 86.8   &      --- & \baseT{>2.1}{19}   \cite{Danevich01}           \\
		      &                  &                        &                       &  $0\nu/2\nu\beta\beta$   &  $2^+_2$ 966.2    &   ---  &   ---            \\
		      &                  &                        &                       &    $0\nu/2\nu\beta\beta$  &  $0^+_1$ 1279.9    &   ---  &   ---           \\
		      &                  &                        &                       &  $0\nu/2\nu\beta\beta$  &  $0^+_2$ 1456.8    &  ---  &   ---          \\ 
\hline
\end{tabular}
\end{center}
($^1$) calculated as described in the text\\
($^2$)  limit derived from the isotopic abundance of daughter nuclide in earth's crust
\medskip
\caption{\label{tab:isotopes} Isotopes and decay modes investigated in this work. Shown is the Gd isotope, the natural isotopic abundance, the daughter isotope, the decay mode, the level state and energy, the Q-value, the theoretical half-life as discussed in the text, and previous experimental constraints. Nuclear data taken from \cite{NuclData}.
}
\end{table*}

\begin{figure*}
  \centering
  \includegraphics[width=0.99\textwidth]{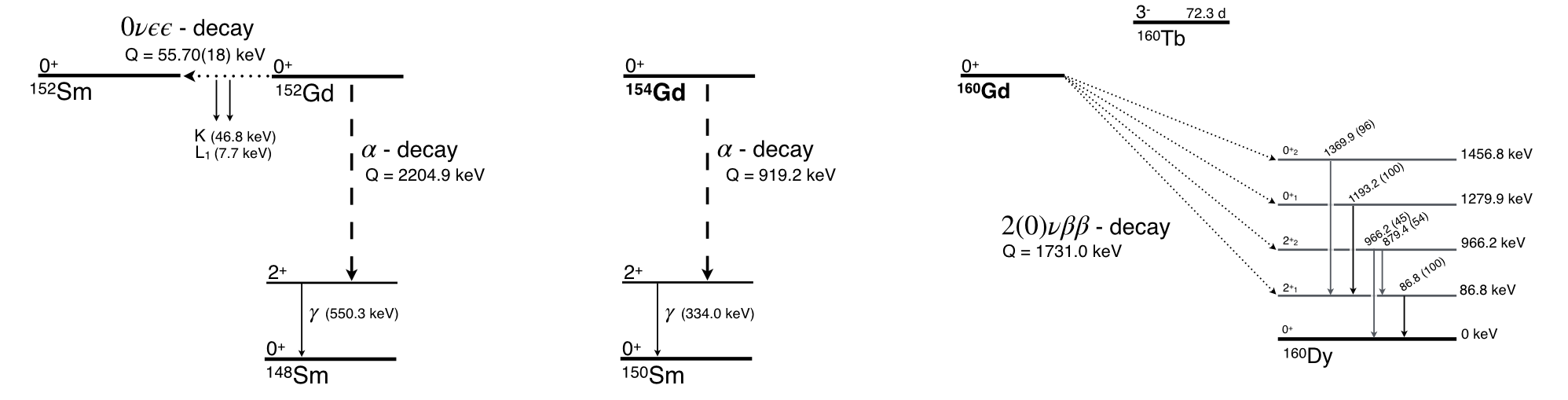}
  %\vspace{5cm}
\caption{Decay schemes of Gd isotopes with double beta decay and alpha decay modes as investigated in this work. \nuc{Gd}{152} can decay via double electron capture and alpha decay with the X-ray and \gray\ as experimental signatures, respectively.  \nuc{Gd}{154} can decay via alpha decay.  \nuc{Gd}{160} can decay via multiple double beta decay modes with the de-excitation \grays\ as the experimental signature. 
Data were taken from \cite{NuclData}.
 }
  \label{pic:decayScheme}
\end{figure*}

We use an ultra-low-background HPGe detector setup in the ``source$\, \neq\, $detector'' approach and all decay modes require a \gray\ or x-ray emissions as an experimental signature.
All investigated Gd isotopes along with their decay modes are listed in \tab \ref{tab:isotopes}. The isotopic abundances of the isotopes, the decay daughters, and investigated excited level as well as the Q-values are listed. Decay schemes are shown in \fig \ref{pic:decayScheme}.\\
The predicted theoretical half-lives for resonant \bbe{0} of \nuc{Gd}{152} ranges between \baseT{6.8}{27}-\baseT{3.8}{30}~yr (assuming $m_{\beta\beta}=100$~meV) and is taken from four independent calculations summarized in \cite{0neeReview}. The only experimental information comes from the isotopic abundance of the daughter nuclide in the earth's crust and is rather weak with  \baseT{>6.0}{8}~yr \cite{Nozzoli18}.
The theoretical half-life for the \nuc{Gd}{160} \bb{2} ground state transition (not investigated in this work) is \baseT{2.95}{21} yr \cite{Delion17} with the experimental limit at \baseT{>1.9}{19}~yr (90\% CL) \cite{Danevich01}. 
Ref.~\cite{Danevich01} also contains the only experimental constraints on the \nuc{Gd}{160} $2^+_1$ excited state transition with $T_{1/2}$ \baseT{>1.3}{21}~yr for the \bb{0} mode and $T_{1/2}$~\baseT{>2.1}{19}~yr for the \bb{2} mode (90\% CL). 

The expected rare alpha decay half-lives into excited states were calculated according to \cite{Poenaru83} taking into account the non-zero momentum transfer for the $0^+_{g.s.} \rightarrow 2^+_1$ transitions. $T_{1/2}$ \baseT{=2}{25}~yr and  \baseT{5}{80}~yr are expected for \nuc{Gd}{152} and \nuc{Gd}{154}, respectively. To our knowledge no experimental constraints are available. 

%%%%%%%%%%%%%%%%%%%%%%%%%%%%%%%%%%%%%%%%%%%%%%%%%%%%%%%%%%%%%
%%%%%%%%%%%%%%%%%%%%%%%%%%%%%%%%%%%%%%%%%%%%%%%%%%%%%%%%%%%%%
%%%%%%%%%%%%%%%%%%%%%%%%%%%%%%%%%%%%%%%%%%%%%%%%%%%%%%%%%%%%%
%\clearpage
%\newpage
\section{Experimental Setup and Sample}

The experimental setup used in this work is located underground in the Gran Sasso National Laboratories of the I.N.F.N.\ (Italy) providing an average overburden of 3600 m w.e.\ against cosmic muons.
The detector consists of a 2.3~kg p-type Ge crystal and has 99\% relative efficiency compared to a 3”$\times$3” NaI(Tl). Its ULB cryostat is constructed of high radiopurity electrolytic copper (\nuc{Th}{228} $ < 12~\mu$Bq/kg, \nuc{Co}{60} $ < 18~\mu$Bq/kg) with only small quantities of other radiopure materials, which, nevertheless, have been screened and cleaned prior to the detector assembly. 
The energy resolution of the spectrometer is 2.0~keV at the 1332~keV \gline\ of \nuc{Co}{60}. The energy-dependent resolution is calibrated using peaks from the internal background. 
The detector is housed in a sample chamber of 5~cm thick electrolytic copper (25$\times$25$\times$35~cm), suitable for accommodating large volume samples. In addition, 20~cm of low radioactivity lead (28~Bq/kg) shields the detector against environmental radiation. To remove radon, the setup is continuously flushed by highly pure nitrogen (stored deep underground for a long time). The entire setup is enclosed in a 1~mm thick steel housing with an interlock for sample insertion and Rn daughter suppression as well as 0.8~mm thick butyl rubber gloves for sample handling.

The Gd-containing sample was purified by the liquid-liquid extraction method adopted from \cite{Polischuk13}, in which the gadolinium-containing compound is pulled from ``solvent 1" to ``solvent 2". The two solvents are immiscible, one is based on water and the other is organic.
In the first stage, the water solution of 800~mL Gd(NO$_3$)$_3$ 
(initial Gd compound) was decomposed to Gd$_2$O$_3$ by preliminary water evaporation followed by its annealing at 900$^\circ$C for 24~h.  Then, the obtained Gd$_2$O$_3$ powder was rinsed with ultra-pure water.  At the next stage, Gd$_2$O$_3$ was dissolved in ultra-pure HCl acid following the reaction
$$
{\rm Gd}_2{\rm O}_3 + 6{\rm HCl}  \rightarrow 2{\rm GdCl}_3 + 3{\rm H}_2{\rm O}\ ,
$$
where reagents were added such that a 20\% acidic solution of gadolinium chloride was achieved. This is solvent 1. As solvent 2, 
trioctylphosphine oxide (TOPO) was used with 0.1~mol/L concentration that purifies the Gd-compound according to the following reaction:
\begin{align*}
\rm GdCl_3(Th, U){\it (aq)} + nTOPO{\it (org)}  \rightarrow  \\
\rm GdCl_3{\it (aq)} + [(Th, U) nTOPO](Cl){\it (org)}\ .
\end{align*}
Further purification occurs during the stage of the gadolinium hydroxide formation and precipitation:
$$
\rm GdCl_3 + 3NH_3 + 3H_2O  \rightarrow  Gd(OH)_3 \downarrow +\, 3NH_4Cl \ .
$$
At the final stage, the obtained amorphous Gd-containing sediment was rinsed several times with ultra-pure water, dried, and annealed in two stages (380$^\circ$C and 600$^\circ$C) for 12 h. The white powder of the final Gd$_2$O$_3$ compound with a mass of 198~g was used for the measurements.

\begin{table}[htbp]
\begin{center}
\begin{tabular}{lcl}
\hline
 element  &   ICP-MS [ppb]  & possible x-rays [keV] (prob.) \\

\hline
  \nuc{Th}{}   &   $<0.1$  \\ 
  \nuc{U}{}   &   $<0.2$  \\ 
  \nuc{K}{}   & $<2$  \\ 
  \nuc{La}{}   & $2$  \\ 
  \nuc{Ce}{}   &  $15$  \\ 
  \nuc{Pr}{}   &  $2$  \\ 
  \nuc{Nd}{}   &  $25$  \\ 
  \nuc{Sm}{}   &   $<200$  & 40.12(47.5\%)	39.52(26.4\%)	\\
  && 45.41(9.15\%)	46.58(3.02\%)	\\
  && 45.29(4.73\%)\\ 
  \nuc{Eu}{}   &   $<50$  & 41.54(47.6\%)	40.90(26.6\%)	\\
  && 47.04(9.21\%)	48.25(3.05\%)	\\
  && 46.91(4.76\%)\\ 
  \nuc{Tb}{}   &   $<1000$  & 44.48(47.5\%)	43.74(26.7\%)	\\
  && 50.38(9.44\%)	51.70(3.15\%)	\\
  && 50.23(4.88\%)\\ 
  \nuc{Dy}{}   &   $50$  & 46.00(47.5\%)	45.21(26.8\%)	\\
  &&52.11(9.58\%)	53.48(3.20\%)	\\
  && 51.95(4.95\%)\\ 
  \nuc{Ho}{}   &   $2$  \\ 
  \nuc{Er}{}   &   $100$  & 49.13(47.5\%)	48.22(27.0\%)	\\
  &&55.67(9.77\%)	57.14(3.28\%)\\
  &&	55.48(5.06\%)\\ 
  \nuc{Tm}{}   &   $<200$  & 50.74(47.4\%)	49.77(27.2\%)	\\
  &&57.51(9.86\%)	59.03(3.32\%)	\\
  && 57.30(5.11\%)\\ 
  \nuc{Lu}{}   &   $<30000$(*)  & 54.07(47.3\%)	52.97(27.3\%)	\\
  && 61.29(10.1\%)	62.93(3.42\%)	\\
  && 61.05(5.21\%)\\ 
\hline
\end{tabular}\\
(*) \nuc{Gd}{157}\nuc{O}{18} interferes with \nuc{Lu}{175} 
\medskip
\caption{\label{tab:bgContent} Contaminations of the Gd$_2$O$_3$ powder sample measured by ICP-MS. 
Uncertainties are about 30\%. The upper limits are given at 68\% C.L.
Possible x-ray lines are quoted for impurities that could interfere with the analysis. Selected are those in the energy range between 35 and 50~keV for impurity concentrations larger than 50~ppb. They are ordered by K-L3, K-L2, K-M3, K-N2N3, and K-M2 transitions and contain the emission probabilities in parentheses.
}
\end{center}
\end{table}

\begin{table}[htbp]
\begin{center}
\begin{tabular}{lcc}
\hline
 nuclide  &   literature [\%] & ICP-MS [\%] \\
\hline
  \nuc{Gd}{152}   &  $0.20\pm 0.01$&  $0.194\pm0.005$   \\ 
  \nuc{Gd}{154}   &  $2.18 \pm 0.03$ &  $2.13\pm0.03$   \\ 
  \nuc{Gd}{155}   &  $14.80\pm 0.12$ &  $14.7\pm0.1$   \\ 
  \nuc{Gd}{156}   &  $20.47\pm 0.09$ & $20.3\pm0.2$  \\ 
  \nuc{Gd}{157}   &  $15.65\pm 0.02$ & $15.7\pm0.1$  \\ 
  \nuc{Gd}{158}   &  $24.84\pm 0.07$ & $25.0\pm0.2$  \\ 
  \nuc{Gd}{160}   &  $21.86\pm 0.19$ & $22.0\pm0.2$  \\ 
\hline
\end{tabular}
\medskip
\caption{\label{tab:isotopicAbundance} Isotopic abundances of gadolinium isotopes in literature \cite{NuclData} and measured by ICP-MS in the investigated Gd$_2$O$_3$ powder sample.
}
\end{center}
\end{table}

ICP-MS measurements were performed to determine impurities concentrations and to assess the isotopic abundance of Gd isotopes. 
A list of analyzed impurities is reported in \tab \ref{tab:bgContent}. The measured concentrations of La, Ce, Pr, Nd, Dy, Ho, and Er are below 100~ppb.
For other tested isotopes only limits could be set. The limit for Lu could only be set at $<30$~ppm due to inference with \nuc{Gd}{157}\nuc{O}{18}.
The last column in \tab \ref{tab:bgContent} also shows possible x-ray emissions from the impurities listed which could interfere with the search. However, in the low-background environment of the measurement we do not expect any significant impurity excitations, and the listed x-ray lines are only shown for completeness and not used in the analysis.  
Isotopic abundances of Gd isotopes from the ICP-MS measurement are reported in \tab \ref{tab:isotopicAbundance} and are consistent with literature values from \cite{NuclData}.

\begin{figure*}
  \centering
  \includegraphics[width=0.99\textwidth]{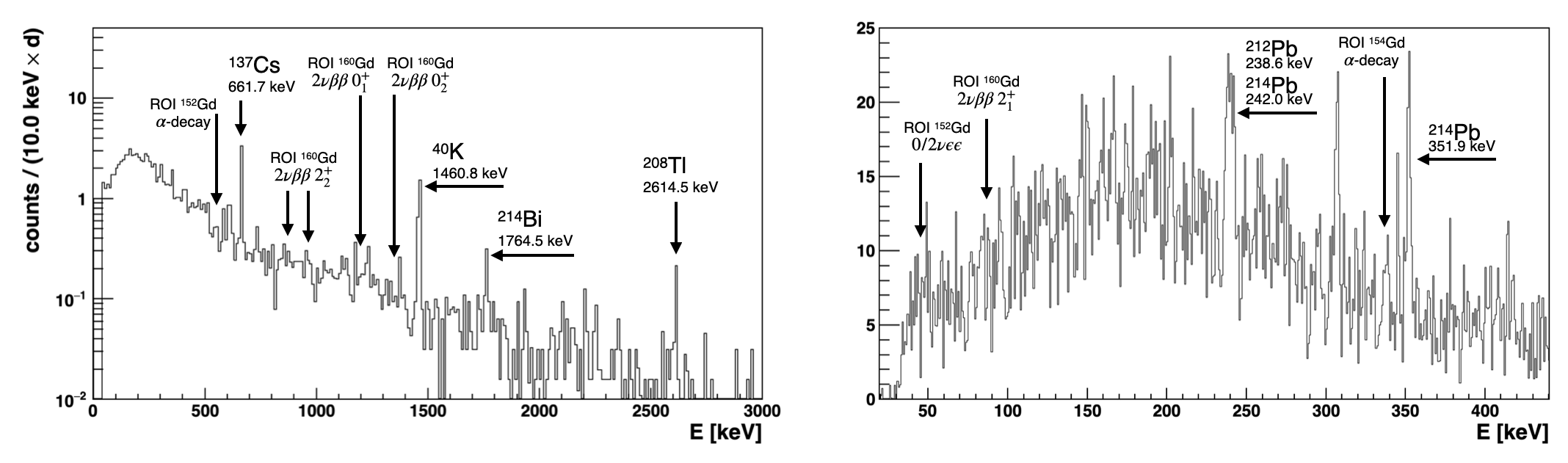}

  %\vspace{5cm}
\caption{Energy spectrum of the 198~g Gd$_2$O$_3$ powder sample measured for 63.8~d with HPGe \gray spectroscopy. Left: coarsely binned full spectrum normalized to counts per day. Right: low energy region in DAQ channels. Highlighted are the regions of interest and prominent background peaks. }
  \label{pic:WideSpec}
\end{figure*}

For the measurement on the ultra-low background HPGe detector, the Gd$_2$O$_3$ powder was sealed in a plastic container and placed onto the endcap. 
Data was taken for 63.8~days. \fig \ref{pic:WideSpec} shows the full spectrum on the left and zooms into the low energy region on the right. Significant peaks and the search regions of interest are labeled.
The dataset was used to determine \gray\ emitting isotopes which are shown in \tab \ref{tab:HPGeBgContent} in comparison to an initial Gd compound based on Gd$_2$(NO$_3$)$_3$. The radiopurity of the Gd$_2$O$_3$ was equal or better for the most prominent nuclides apart from \nuc{Cs}{137}.

\begin{table}[htbp]
\begin{center}
\begin{tabular}{rrcc}
\hline
 & &  \multicolumn{2}{c}{activity [mBq/kg]} \\
 \hline
  chain & nuclide      &  Gd$_2$O$_3$  & Gd$_2$(NO$_3$)$_3$   \\
\hline
%Th232
\nuc{Th}{232} & \nuc{Ra}{228}   & $1.0\pm0.5$  & $0.9\pm0.3$    \\ 
 &  \nuc{Th}{228}   & $2.0\pm0.4$  & $1.1\pm0.2$   \\

% U238
\nuc{U}{238} &   \nuc{Ra}{226}   & $3.2\pm0.5$  & $6.4\pm0.5$   \\ 
 & \nuc{Th}{234}   & $< 110$  & $< 210$   \\ 
&  \nuc{Pa}{234m}   & $<41$    & $<18$   \\ 

 \nuc{U}{235} & \nuc{U}{235}     & $< 2.0$  & $< 0.9$  \\ 
\\
% other

 &  \nuc{K}{40}       & $3.3\pm0.5$  & $15\pm3$   \\ 
  &\nuc{Cs}{137}   & $4.2\pm0.6$  & $0.5\pm0.1$   \\ 

 & \nuc{Eu}{152}     & $< 1.5$  & $< 0.6$  \\ 
  &\nuc{Eu}{154}     & $< 1.6$  & $< 0.7$  \\ 

\hline
\end{tabular}
\medskip
\caption{\label{tab:HPGeBgContent} Radioactive contamination of the Gd$_2$O$_3$ powder sample used for this work and a previously studied sample of a Gd$_2$(NO$_3$)$_3$ water solution. The upper limits are given at 90\% C.L., and the uncertainties of the measured activities at 68\% C.L.
}
\end{center}
\end{table}

%%%%%%%%%%%%%%%%%%%%%%%%%%%%%%%%%%%%%%%%%%%%%%%%%%%%%%%%%%%%%
%%%%%%%%%%%%%%%%%%%%%%%%%%%%%%%%%%%%%%%%%%%%%%%%%%%%%%%%%%%%%
%%%%%%%%%%%%%%%%%%%%%%%%%%%%%%%%%%%%%%%%%%%%%%%%%%%%%%%%%%%%%
\section{Analysis}

\begin{table}[t]
\begin{center}
\begin{tabular}{cccc}
\hline
\gline & emission  & detection  & total  \\
  \ [keV]     & prob.  & prob.  & prob.  \\
\hline
39.5  & 0.227   &  \baseT{0.8\pm0.2}{-8}  &   \baseT{1.8\pm0.2}{-9}  \\
40.1  & 0.408   &  \baseT{1.5\pm0.2}{-8}  &   \baseT{6.2\pm0.8}{-9}  \\
45.3  & 0.041   &  \baseT{1.0\pm0.1}{-6}  &   \baseT{4.1\pm0.4}{-8}  \\
45.4  & 0.079   &  \baseT{1.0\pm0.1}{-6}  &   \baseT{7.9\pm0.8}{-8}  \\
46.5  & 0.026   &  \baseT{2.3\pm0.2}{-6}  &   \baseT{6.0\pm0.6}{-8}  \\
\hline
\end{tabular}
\medskip
\caption{\label{tab:Expsignature} Emission, detection, and total probabilities for the experimental signature of x-rays from resonant \bbe{0} decay of \nuc{Gd}{152}. }
 \end{center}
\end{table}

The analysis is based on peak searches for the de-excitation \grays\ of each independent decay mode using the Bayesian framework BAT (Bayesian Analysis Toolkit) \cite{Caldwell:2009kh}. 
The likelihood is defined as the product of the Poisson probabilities over each bin. The expectation in each bin $i$ is the sum of the signal $S_i$ and background $B_i$ expectation. 

$S_i$ is the integral of the Gaussian peak shape in the bin given the total signal peak counts $s$
\begin{eqnarray}
S_i&=&
 \int_{\Delta E_{i}}  \frac{s}{\sqrt{2\pi}\sigma_E} 
\cdot \exp{\left(-\frac{(E-E_0)^2}{2\sigma_E^2}\right)} dE\ , \label{eq:Si}
\end{eqnarray}

where $\Delta E_{i}$ is the bin width, $\sigma_E$ the energy resolution and $E_0$ the \gline\ energy as the mean of the Gaussian.

$B_i$, the background expectation 
\begin{eqnarray}
B_i = \int_{\Delta E_{i}}&& b + c\left( E-E_0 \right) \\
+&&   \sum \limits_{l} \left[\frac{b_{l}}{\sqrt{2\pi}\sigma_{l}} \cdot \exp{\left(-\frac{(E-E_{l})^2}{2\sigma_{l}^2}\right)}\right] dE \nonumber
\label{eq:Bi}
\end{eqnarray}

which is implemented as a linear function (parameters $b$ and $c$) and $l$ Gaussian background peaks in the fit window, depending on the decay mode.

The signal counts are connected with the half-life $T_{1/2}$ of the decay mode as
\begin{eqnarray}
\label{eq:HLtoCounts}
s =
\ln{2} \cdot  \frac{1}{T_{1/2}} \cdot \epsilon \cdot N_A \cdot T \cdot m \cdot f  \cdot \frac{1}{M}\ ,
\end{eqnarray}

where $\epsilon$ is the full energy peak detection efficiency,
$N_A$ is the Avogadro constant,
$T$ is the live-time (63.8~d), 
$m$ is the mass of Gd in the sample (171.8~g), 
$f$ is the isotopic fraction of the respective Gd isotope, 
and $M$ its the molar mass of natural Gd (157.25).

\begin{table*}[htbp]
\begin{center}
\begin{tabular}{llllll}
\hline
nuclide (decay) & daughter (level)  &    \glines  &  det.\ eff.\ $\epsilon$   & $\sigma_{\rm res}$ &  T$_{1/2}$ (90\% C.I.) \\
                         &  ($J^\pi $ keV)  &     [keV]                    &  [\%]               &  [keV]                      &    [yr] \\
\hline

\nuc{Gd}{152} (\bbe{0}) & \nuc{Sm}{152} ($0^+_{\rm g.s.})$   &   39.5-46.5    &  \baseTsolo{-6} - \baseTsolo{-4}   &   0.60  &   \baseT{>4.2}{12}               \\

\nuc{Gd}{160} $(0\nu/2\nu\beta\beta)$ & \nuc{Dy}{160} ($2^+_1 86.8)$   &   86.8    & 0.088 &   0.66  &   \baseT{>1.8}{18}               \\
\nuc{Gd}{160} $(0\nu/2\nu\beta\beta)$  & \nuc{Dy}{160} ($2^+_2 966.2)$   &   879.4    & 1.8   &   1.02  &    combined fit              \\
 										  &    &   966.2    & 1.4   &   1.04  &     \baseT{>9.7}{19}              \\
\nuc{Gd}{160} $(0\nu/2\nu\beta\beta)$ & \nuc{Dy}{160} ($0^+_1 1279.9)$   &   1193.2    & 2.8   &   1.09  &   \baseT{>8.2}{19}               \\
\nuc{Gd}{160} $(0\nu/2\nu\beta\beta)$  & \nuc{Dy}{160} ($0^+_2 1456.8)$   &   1369.9    & 2.6   &   1.12  &   \baseT{>5.0}{19}               \\

\hline
\nuc{Gd}{152} ($\alpha$) & \nuc{Sm}{148} ($2^+_1 550.3)$   &   550.3    & 4.0   &   0.92  &   \baseT{>3.4}{17}               \\
\nuc{Gd}{154} ($\alpha$) & \nuc{Sm}{150} ($2^+_1 333.9)$   &   333.9    & 4.7   &   0.83  &   \baseT{>9.6}{18}               \\

\hline
\end{tabular}
\medskip
\caption{\label{tab:HLimits} Lower half-life limits on different decay modes that can occur in gadolinium isotopes. Columns 3-5 show the \glines\ used in the fit together with their detection efficiency and energy resolution. In the case of multiple \glines\ a combined fit is used for setting a half-life limit.  }
\end{center}
\end{table*}

Each free parameter is assigned a prior probability. The inverse half-life $(T_{1/2})^{-1}$ and linear background parameters have flat prior distributions. The priors for energy resolution, peak position, and detection efficiencies are Gaussian distributions centered around the respective mean values of these parameters, with the width determined by the parameter uncertainty. This approach naturally accounts for systematic uncertainties in the fitting results.

The peak position uncertainty is fixed at 0.1~keV, while the energy scale and resolution are obtained from standard calibration spectra with an estimated uncertainty of 5\%. The full energy peak detection efficiencies are determined using MaGe Monte-Carlo simulations \cite{bos11} and reported with associated uncertainties in \tab \ref{tab:Expsignature}. Systematic uncertainties regarding the measured sample mass and the isotopic fraction in the sample (as shown in \tab \ref{tab:isotopicAbundance}) are relatively small in comparison to the detection efficiency uncertainty and therefore are not considered.

Background \glines\ above 1\% emission probability from the \nuc{U}{238} and \nuc{Th}{232} decay chains are included in all fit windows.
For the 35-56~keV fit window of the \nuc{Gd}{152} \bbe{0} search this includes 
39.9~keV (1.1\%, \nuc{Bi}{212}), 
46.5~keV (4.3\%, \nuc{Pb}{210}), and  
53.2~keV (1.1\%, \nuc{Pb}{214}). The emission probabilities and isotopes are in parentheses.

\begin{figure*}
  \centering
  \includegraphics[width=0.49\textwidth]{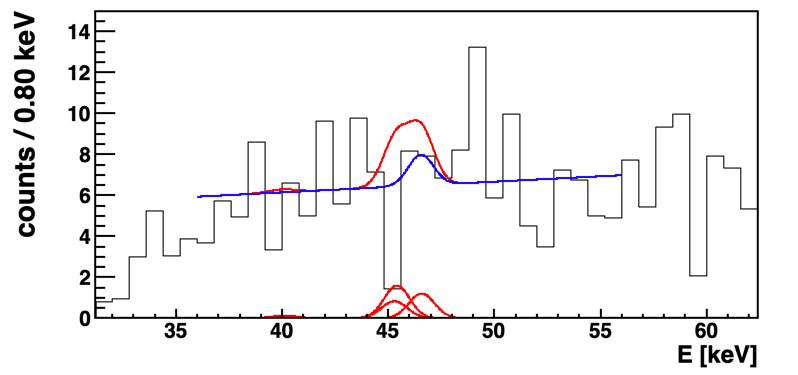}
  \includegraphics[width=0.49\textwidth]{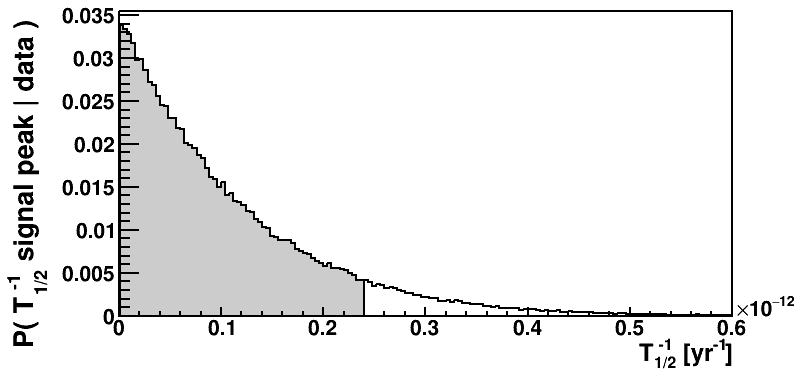}

  %\vspace{5cm}
\caption{Fit result of \nuc{Gd}{152} \bbe{0} transition. Left: The blue curve is the best fit for the data shown in the histogram. The red curve is the 90\% CI determined from the fit shown as the combined p.d.f.\ as well as individual peak components at the bottom. Right: Marginalized posterior of the inverse half-life. The 90\% quantile is shaded and used for limit setting.}
  \label{pic:FitResult}
\end{figure*}

The likelihoods and prior probabilities are used by BAT to obtain the full posterior probability distribution using Markov Chain Monte Carlo. The multi-dimensional posterior is then marginalized to $(T_{1/2})^{-1}$ as the parameter of interest. The results are shown in \fig \ref{pic:FitResult} using the \nuc{Gd}{152} \bbe{0} as an example. The left plot shows the spectrum and model function. The blue curve denotes the best fit and the red curve shows the excluded peak strength at 90\% credibility. At the bottom, all x-ray lines are shown individually. The right plot shows the marginalized inverse half-life. The best fit for this transition is at zero. The 90\% quantile is used to set the 90\% credibility limit at \baseT{4.2}{12}~yr.

The best-fit values for all other decay modes are also consistent with zero signal counts.
The 90\% credibility half-life limits are shown for all investigated decay modes in \tab \ref{tab:HLimits}. They range between \baseTsolo{17} and \baseTsolo{20}~yr. Also shown are the \gline\ energies, the full energy peak detection efficiency, and the resolution for the \glines\ used in the analyses. 
For the  \nuc{Gd}{160} $0\nu / 2\nu \beta\beta$ $2^+_2$ decay mode, two \glines\ are used, each having its own fit window, likelihood, and free parameters but sharing the same half-life parameter in \eq \ref{eq:HLtoCounts}.

%%%%%%%%%%%%%%%%%%%%%%%%%%%%%%%%%%%%%%%%%%%%%%%%%%%%%%%%%%%%%
%%%%%%%%%%%%%%%%%%%%%%%%%%%%%%%%%%%%%%%%%%%%%%%%%%%%%%%%%%%%%
%%%%%%%%%%%%%%%%%%%%%%%%%%%%%%%%%%%%%%%%%%%%%%%%%%%%%%%%%%%%%
\section{Discussion and Conclusions}

The main result of this search is the first direct detection limits of the resonant neutrinoless double electron capture of \nuc{Gd}{152}.
Existing constraints from geological measurements of daughter abundances are improved by about 4 orders of magnitude. 
In addition, $0\nu / 2\nu \beta\beta$ decay constraints on the first three excited state transitions of \nuc{Gd}{160} are obtained. Those are the first experimental results for the $2^+_2$ and $0^+_1$ transitions. Furthermore, rare alpha decays into the first excited states were tested in \nuc{Gd}{152} and \nuc{Gd}{154}.  
None of the investigated decay modes yield a significant signal and 90\% credibility limits are set with a Bayesian analysis taking into account the dominant systematic uncertainties. 

The resonant \bbe{0} of \nuc{Gd}{152} could be excluded to be slower than \baseT{4.2}{12}~yr (90\% CI) using a multi-peak search of the dominant low-energy x-rays atomic shell relaxations.  
This is still many orders of magnitude away from theoretical predictions of \baseT{6.8}{27}$-$\baseT{3.8}{30}~yr for $m_{\beta\beta}=100$~meV. 

Current large-scale \bbm{0} decay experiments exclude $m_{\beta\beta}$ on the order of 100~meV with experimentally achieved half-life limits of $10^{26}$~yr. In general, it appears that the resonant neutrinoless double electron capture in \nuc{Gd}{152} is about 2 orders of magnitude less sensitive to lepton number violation than e.g.\ \nuc{Ge}{76}. However, the comparison between \bbm{0} and resonant \bbe{0} decay as well as between different isotopes is only valid assuming light Majorana neutrino exchange. 
In addition, small changes in the energy split between \nuc{Gd}{152} and its daughter \nuc{Sm}{152} (current best value at $\Delta = 0.91(18)$~keV), can enhance the decay rate significantly, i.e.\ if more resonance is realized in nature. Hence it is worthwhile pursuing future experiments with \nuc{Gd}{152}. 

Experimental improvements of \bbe{0} searches in \nuc{Gd}{152} are difficult in the source$\neq$detector experimental approach. This is due to the low-energy nature of the signal and the multitude of possible x-ray emissions. An immediate improvement is the use of a low-background low-energy detector system e.g.\ an n-type HPGe detector without a dead layer. This is likely to improve the sensitivity from this search by multiple orders of magnitude by increasing the detection efficiency of \baseTsolo{-6}-\baseTsolo{-4} to the percent level. This could be enhanced by placing a thin sample around the HPGe crystal inside the cryostat as e.g.\ done in \cite{Nagorny21}. However, this approach is ultimately limited by the finite sample mass and geometry from which low-energy x-rays can escape. 

Apart from specialized thin-sample large-area detectors, the source$=$detector experimental approach, where the isotopes of interest are embedded into the detector material, would allow larger scalability.
Examples are Gd$_2$SiO$_5$(undoped) or Gd$_2$SiO$_5$(Ce) scintillating crystals operated as cryogenic scintillating bolometers. The advantages are a significant enhancement of the detection efficiency (close to 100\%), especially for low-energy \grays. Moreover, it would allow much more sample mass to be probed. Alpha decays would be detected directly, making it possible to observe ground-state transitions. Pulse-shape discrimination of alpha and beta/gamma events would allow a near background-free search for alpha decay modes.

The higher energy signature of the $0\nu/2\nu\beta\beta$ in \nuc{Gd}{160} (Q-value  1731.0~keV), together with a high natural abundance (21.86\%), makes \nuc{Gd}{160} an interesting and practical isotope. 
Many low-background experiments employ gadolinium for neutron detection which can result in significant double beta decay constraints for \nuc{Gd}{160} from peripheral analyses in these experiments.
The Super-Kamiokande experiment was recently loaded with 13~tons of Gd$_2$(SO$_4$)$_3\cdot 8$H$_2$O, making this the largest deployment of a double beta decay isotope in a low-background environment. Unfortunately, the decay energy shared between the two electrons will likely not be high enough to be reasonably used in a water-Cherenkov detector. 
Gadolinium-loaded scintillators are the most promising approach. Some large-scale dark matter experiments, such as LZ, utilize gadolinium-loaded scintillators as neutron vetoes. In the near future, the best constraints on $0\nu/2\nu\beta\beta$ in \nuc{Gd}{160} will likely come from peripheral analyses in those experiments.

%%%%%%%%%%%%%%%%%%%%%%%%%%%%%%%%%%%%%%%%%%%%%%%%%%%%%%%%%%%%%
%%%%%%%%%%%%%%%%%%%%%%%%%%%%%%%%%%%%%%%%%%%%%%%%%%%%%%%%%%%%%
%%%%%%%%%%%%%%%%%%%%%%%%%%%%%%%%%%%%%%%%%%%%%%%%%%%%%%%%%%%%%
\begin{acknowledgements}

The authors would like to thank Dr.\ Roman Boiko for his valuable recommendations related to the gadolinium powder purification, fruitful discussions, and long-term collaboration. We also thank the staff at the chemistry lab at LNGS for their continuous support, guidance, and assistance, as well as for the use of their equipment.

\end{acknowledgements}

\bibliographystyle{spphys}       % APS-like style for physics


\end{document}